# Photophysical comparison of liquid and mechanically exfoliated WS$_2$ monolayers

Zhaojun Li[1,2], Farnia Rashvand[3], Hope Bretscher[1], Beata M. Szydłowska[3], James Xiao[1], Claudia Backes[3,4], Akshay Rao[1*]

[1]Cavendish Laboratory, University of Cambridge, JJ Thomson Avenue, CB3 0HE, Cambridge, United Kingdom

[2]Molecular and Condensed Matter Physics, Department of Physics and Astronomy, Uppsala University, 75120 Uppsala, Sweden

[3]Institute for Physical Chemistry, Ruprecht-Karls-Universität Heidelberg, Im Neuenheimer Feld 253, 69120 Heidelberg, Germany

[4]Current address: Physical Chemistry of Nanomaterials, University of Kassel, Heinrich-Plett-Str.40, D-34132 Kassel, Germany

[*]Corresponding author

Contact e-mail: ar525@cam.ac.uk

**Abstract**

Semiconducting transition metal dichalcogenides (TMDs) are desired as active materials in optoelectronic devices due to their strong excitonic effects. They can be exfoliated from their parent layered materials with low-cost and for mass production *via* a liquid exfoliation method. However, the device application of TMDs prepared by liquid phase exfoliation is limited by their poor photoluminescence quantum efficiencies (PLQE). It is crucial to understand the reason to low PLQE for their practical device development. Here we evaluate the quality of monolayer-enriched liquid phase exfoliated (LPE) WS$_2$ dispersions by systematically investigating their optical and photophysical properties and contrasting with mechanically exfoliated (ME) WS$_2$ monolayers. An in-depth understanding of the exciton dynamics is gained with ultrafast pump-probe measurements. We reveal that the energy transfer between monolayer and few-layers in LPE WS$_2$ dispersions is a substantial reason for their quenched PL. In addition, we show that LPE WS$_2$ is promising to build high performance optoelectronic devices with excellent optical quality.

**Introduction**

The study of transition metal dichalcogenides (TMDs) has become a vibrant area in nanomaterial science.[1,2] Exfoliated TMDs have been widely used in the field of optoelectronics due to their excellent light absorptivity and semiconducting performance.[3–6] To utilize TMD materials, achieving more scalable techniques is critically important, and considerable effort has been devoted to the development of cost-effective mass production methods.[7,8] Although mechanical exfoliation produces the highest quality materials, its application is limited by extremely low and uncontrollable yield.[9,10] In contrast, liquid exfoliation yields atomically thin TMD flakes in a liquid medium in large quantities at moderate cost.[11,12] The simplest way to produce nanosheets suspended in liquid is termed liquid phase exfoliation (LPE) which relies on immersing the bulk materials into suitable solvents or aqueous surfactant solution and apply high energy, *e.g.* sonication, to achieve exfoliation accompanied with tearing.[13,14] When appropriately chosen, the solvent or surfactant suppresses reaggregation in the liquid.[15] This approach is widely applicable to a range of materials and takes advantage of well-established print production processes for new generation of device fabrication.[16,17] Nevertheless, liquid-exfoliated (LE) TMD nanosheets have been rarely used in optoelectronics, which is often attributed to their poor quality, such as defects, impurities, nonuniformity and small size, resulting in low photoluminescence quantum efficiencies (PLQE).[18] Hence the development of high PLQE LE TMD monolayers is of fundamental importance toward the practical implementation of optoelectronic devices with TMDs. Nowadays, the quality of the LPE TMDs is improving with the continual development of the liquid phase exfoliation methodology and subsequent size selection.[19] Largely defect-free monolayer enriched TMD dispersions with narrow line width photoluminescence (PL), which are similar to that of mechanically-exfoliated (ME) TMDs, were demonstrated.[20,21] However, the PLQE of LPE TMD dispersions remains low and exciton dynamics of TMD dispersions is only little explored.[22] In particular, most reports focus on ensembles with low monolayer content produced either from LPE or colloidal synthesis.[23–26] Recently, we achieved high quality WS$_2$ samples by mechanical exfoliation with bis(trifluoromethane)sulfonimide lithium salt (Li-TFSI) surface treatment, showing superior PL. In this work, we evaluate the quality of monolayer enriched WS$_2$ dispersion produced from liquid phase exfoliation by conducting a systematic study on the optical and photophysical properties of different LPE WS$_2$ dispersions and ME WS$_2$ monolayer samples. The LPE WS$_2$ dispersions can achieve PL with narrow linewidth and almost no Stoke shift compared to their absorption spectra. In addition, we use ultrafast pump-probe spectroscopy to study the exciton dynamics following photoexcitation in these WS$_2$ samples. We reveal that the energy transfer between monolayers and few-layers in LPE WS$_2$ dispersion samples can be a major factor for their quenched PL.

**Results and Discussion**

The LPE WS$_2$ dispersion samples used in this study were prepared as shown in Figure 1. The liquid-suspended WS$_2$ nanosheets are generated with the aid of dip sonication and stabilized against reaggregation by the surfactant sodium cholate in water. A size selection process is followed since the as-produced dispersion is highly polydisperse displaying a low monolayer content. Size selection is achieved by liquid cascade centrifugation (LCC) with subsequently increasing rotational speeds.[27] Heavier and multilayer nanosheets are removed in each step of the LCC process, resulting in more and more monolayer-enriched supernatants. Two size-selected nanosheet distributions are collected as sediments after 10k *g* and 30k *g* centrifugation, hereafter labelled as 5-10k *g* WS$_2$/H$_2$O sample and 10-30k *g* WS$_2$/H$_2$O sample, respectively. Isopropyl alcohol (IPA) is also known to give stable dispersions, however monolayer enrichment has not yet been demonstrated. Hence, a 10-30k *g* WS$_2$/IPA sample is also prepared in comparison by replacing the water/surfactant in the 10-30k *g* WS$_2$/H$_2$O sample with IPA through a centrifugation procedure. The monolayers obtained in this way are around 50 nm as characterized by transmission electron microscopy (TEM) (Figure S1a), which is similar to what was obtained from previous work.[20] Since mechanical exfoliation renders high quality TMD monolayers, ME WS$_2$ samples are also prepared as reference to LPE WS$_2$ samples. Large monolayer WS$_2$ samples (~ 200 µm) prepared on quartz substrates with mechanical exfoliation are identified by optical microscopy (Figure S1b). As shown in Figure S1c, both LPE and ME WS$_2$ samples on Si/SiO$_2$ substrates are characterized by Raman spectroscopy (excitation wavelength 532 nm), confirming the monolayer with characteristic Raman modes of monolayer WS$_2$ (e.g., the 2LA(M) at 354 cm$^{-1}$)[28,29].

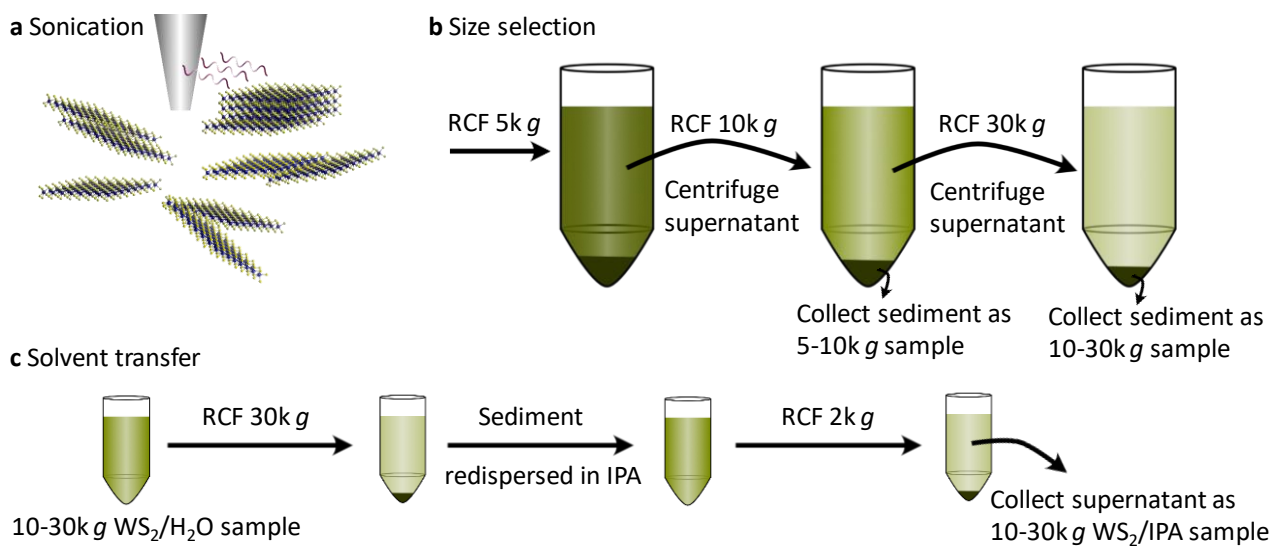

Figure 1. Illustration of liquid exfoliation, size selection and solvent transfer. **a** Schematic of the tip sonication. **b** Schematic of the liquid cascade centrifugation. Relative centrifugal forces of (RCF) 5k $g$, 10k $g$, and 30k $g$ are used. The supernatant after each step is transferred to another centrifugation at higher centrifugal acceleration, while the sediments are collected. **c** Schematic of the dispersion solvent exchange, transferring from H$_2$O to IPA. 30k $g$ RCF is used to spin down the WS$_2$ nanosheets as a pellet allowing for a solvent exchange. 2k $g$ RCF is used to remove aggregated nanosheets as sediments.

The optical absorption properties of LPE 5-10k $g$ WS$_2$/H$_2$O, 10-30k $g$ WS$_2$/H$_2$O, and 10-30k $g$ WS$_2$/IPA dispersions are characterized via UV-Vis extinction spectroscopy. The spectra depicted in Figure 2a and Figure S2a are normalized to the local minimum 290 nm, since the extinction coefficient at 290 nm is widely independent of nanosheet thickness and length.[30] The absorption spectra are dominated by excitonic features. The fundamental A-exciton ($E_A^{ML}$) for all LPE WS$_2$ dispersions is analysed in more detail using the second derivative of the extinction spectra (Figure 2b and Figure S2b). Due to the previously identified exponential blueshift of the A-exciton with decreasing layer number, two components are visible in the second derivative attributed to the A-exciton of the monolayer ($E_A^{ML}$) and the unresolvable sum of few-layers ($E_A^{FL}$).[20,21,31] WS$_2$ dispersions in H$_2$O show $E_A^{ML}$ at 2.029 eV (611 nm), while WS$_2$ dispersions in IPA present slightly redshifted $E_A^{ML}$ at 2.019 eV (614 nm), which may be attributed to solvatochromism and difference of dielectric disorder.[32] Since the contributions to the A-exciton absorbance of mono- and few-layer WS$_2$ nanosheets are differentiated, the monolayer content is estimated from the second derivative of A-exciton absorbance peak according to the previously reported method (described in SI).[20] There is a clear increasing monolayer volume fraction ($V_f$) in water dispersions with increasing RCF, which is 17% for the 5-10k $g$ WS$_2$/H$_2$O sample and 78% for the 10-30k $g$ WS$_2$/H$_2$O sample, respectively. On the other hand, the 10-30k $g$ WS$_2$/IPA sample shows a moderate V$_f$ at around 35% suggesting that some aggregation occurred during the solvent transfer. In the following, we focus on the optical and photophysical properties of the monolayer-enriched LPE 10-30k $g$ WS$_2$/H$_2$O dispersion and 10-30k $g$ WS$_2$/IPA dispersion samples.

To evaluate the quality of LPE WS$_2$ samples, we start by comparing the steady-state PL profiles of LPE and ME WS$_2$ samples. The PL of ME WS$_2$ monolayers are measured with a confocal PL setup, while the LPE samples are measured as dispersions. As shown in Figure 2c, the PL position of ME pristine monolayer WS$_2$ sample ($E_A^{ML}$, PL) is around 1.981 eV with a full width at half maximum (FWHM) value around 44 meV, indicating the emission mainly stems from a dominating contribution of trions in the sample.[10] After Li-TFSI treatment, the PL of ME monolayer WS$_2$ is greatly enhanced and the peak position blueshifts accompanied by a more uniform emission profile due to the suppression of trions and defects, as shown in scatter plots of the peak PL counts versus emission peak position acquired from PL spatial maps (Fig. S3). In addition, the Li-TFSI treated WS$_2$ sample exhibits a narrower FWHM around 10 meV. The PL Stokes shift of the ME Li-TFSI treated WS$_2$ sample is primarily related to strain.[33] This is in good agreement with our previous work showing that Li-TFSI treatment can minimize trap and trion states resulting in intrinsic monolayer properties.[9,34] The PL positions of both the LPE 10-30k $g$ WS$_2$/H$_2$O dispersion and the 10-30k $g$ WS$_2$/IPA dispersion coincide with that of the monolayer A-exciton absorbance with almost no Stokes shift, suggesting a high optical quality of the samples with near intrinsic properties, Table 1. However, the monolayer enriched LPE WS$_2$ dispersions show extremely low PLQE, less than 0.1% as it is too low to determine accurately with our setup. The FWHM value is around 19 meV for both LPE WS$_2$ samples, which is wider compared to that of ME Li-TFSI treated WS$_2$ samples. This may be ascribed to polydispersity induced defect-related broadening of the exciton resonances.[35] Also, the long PL tail below the bandgap of WS$_2$ in the LPE 10-30k g WS$_2$/IPA dispersion is attributed to the larger portion of few layers caused by aggregation.

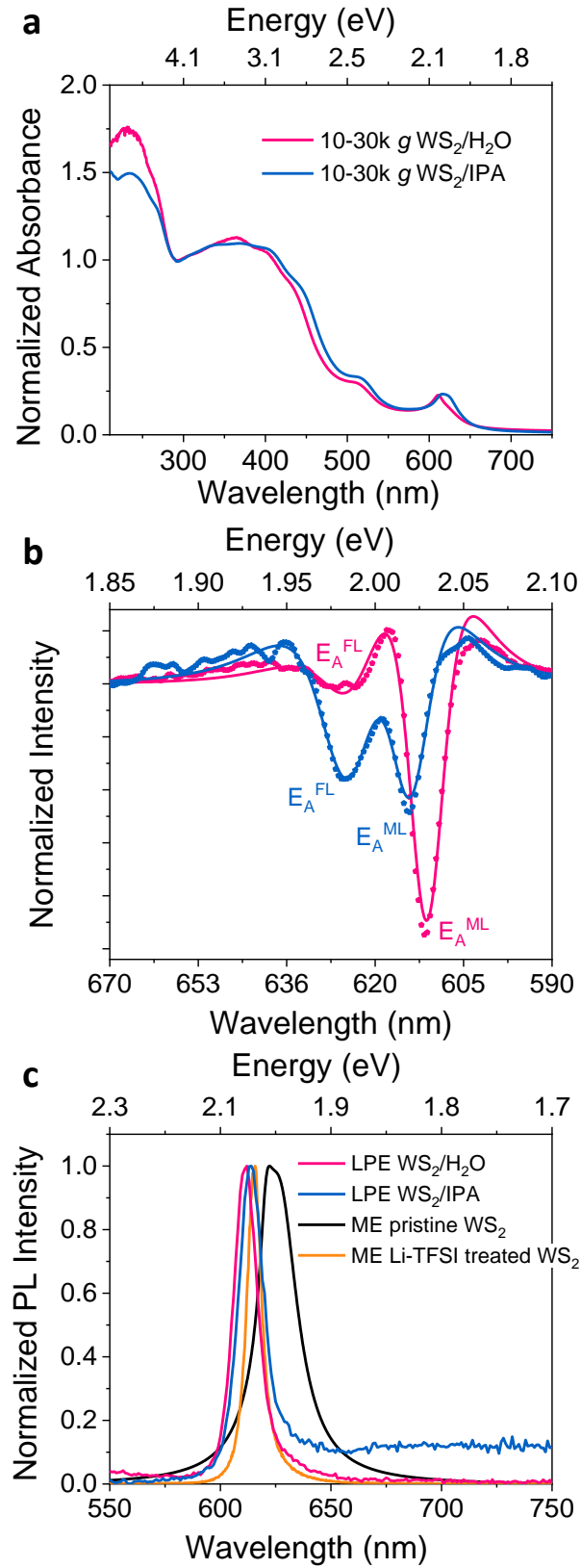

Figure 2. **a** Extinction spectra of the 10-30k *g* WS$_2$/H$_2$O and 10-30k *g* WS$_2$/IPA dispersion samples (normalized to 290 nm). **b** Second derivatives of the A-exciton obtained after smoothing the spectrum with the Lowess method. The spectra are fitted to the second derivative of two Lorentzians (described

in the SI). The positions of monolayer A-exciton ($E_A^{ML}$) and few-layer A-exciton ($E_A^{FL}$) are marked in the figure. **c** Normalized PL spectra of the liquid-exfoliated 10-30k *g* WS$_2$/H$_2$O and 10-30k *g* WS$_2$/IPA dispersion samples as well as mechanically exfoliated pristine and Li-TFSI treated WS$_2$ samples.

Table 1. Summary of monolayer volume fraction ($V_f$), the position of monolayer A-exciton peak ($E_A^{ML}$), center and full width at half-maximum (FWHM) of PL spectra, as well as the position and average exciton lifetime $<\tau>$ of A-exciton ground state bleach (GSB) from pump-probe measurements.

| Sample | ML $V_f$ (%) | $E_A^{ML}$ (Abs, eV) | $E_A^{ML}$ (PL, eV) | FWHM (PL, meV) | $E_A^{ML}$ (GSB, eV) | $<\tau>$ ($A_{ML}$-exciton GSB, ps) | $<\tau>$ ($A_{FL}$-exciton GSB, ps) |
|---|---|---|---|---|---|---|---|
| LPE 10-30k *g* WS$_2$/H$_2$O | 78 | 2.029 | 2.029 | 19 | 2.029 | 481 | 325 |
| LPE 10-30k *g* WS$_2$/IPA | 35 | 2.019 | 2.019 | 19 | 2.019 | 231 | 759 |
| ME pristine WS$_2$ | / | / | 1.981 | 44 | 2.006 | 6 | / |
| ME Li-TFSI treated WS$_2$ | / | / | 2.013 | 10 | 2.013 | 90 | / |

In order to investigate the reason for the low PLQE of LPE WS$_2$ dispersions, we conducted ultrafast pump-probe spectroscopy to explore the exciton dynamics of the LPE WS$_2$ dispersions and the ME WS$_2$ monolayer samples. Upon excitation, the state filling of the A-exciton leads to a reduction in the ground-state absorption, which is referred to as the A-exciton ground-state bleach (GSB). We record the differential transmission ($\Delta T/T$) of a white light probe beam as a function of time after photoexcitation by a pulsed laser. The full pump-probe spectra of the LPE 10-30k *g* WS$_2$/H$_2$O dispersion, the WS$_2$/IPA dispersion, the ME pristine WS$_2$, and the Li-TFSI treated WS$_2$ samples excited at around the A-exciton resonance 610 nm (2.033 eV) with 2.63 nJ/pulse are shown in Figure S4. Dynamic screening of Coulomb interaction gives rise to either a comparatively small red-shift or blue-shift of the A-exciton resonance depending on the exciton density.[36,37] As shown in Figure 3 and summarized in Table 1, the monolayer A-exciton GSB maximum ($E_A^{ML}$, GSB) is located at 611 nm (2.029 eV), 614 nm (2.019 eV) and 616 nm (2.013 eV) for the LPE WS$_2$/H$_2$O dispersion, the WS$_2$/IPA dispersion and the ME Li-TFSI treated WS$_2$ sample, respectively. This coincides with $E_A^{ML}$ (PL), confirming the PL of LPE WS$_2$ dispersions and ME Li-TFSI treated WS$_2$ sample stems from neutral exciton emission. While $E_A^{ML}$ (GSB) is detected at 618 nm (2.006 eV) for the ME pristine WS$_2$ monolayer, which is redshifted compared to $E_A^{ML}$ (PL). This is in good agreement with our interpretation that PL of the ME pristine WS$_2$ monolayer is dominated by trion emission. In contrast to the ME monolayers, LPE samples also display a positive feature at around 650 – 690 nm which we assign to the few-layer A-exciton GSB ($E_A^{FL}$, GSB). The few-layer signal is more prominent in the LPE WS$_2$/IPA sample than that in the 5-10k $g$ and 10-30k $g$ WS$_2$/H$_2$O samples (Figure S4 and S5), even though the monolayer content in the WS$_2$/IPA dispersion is larger than that in the 5-10k $g$ sample. This suggests that this feature is a signature of aggregated nanosheets that is increased in content during the solvent exchange process (Detailed discussion in SI).

The normalized kinetics taken at the $A_{ML}$-exciton GSB and $A_{FL}$-exciton GSB are shown in Figure 3c and the averaged decay lifetimes ($<\tau>$) are summarized in Table 1 while the fitting results are exhibited in Table S2. For the ME pristine monolayer WS$_2$ sample, photogenerated excitons decay primarily through nonradiative exciton-exciton annihilation (EEA) on a few picoseconds time scale.[38,39] After Li-TFSI treatment, $<\tau>$ increases to tens of picoseconds, which is ascribed to radiative recombination. On the other hand, the $A_{ML}$-exciton GSB and $A_{FL}$-exciton GSB for the LPE 10-30k $g$ WS$_2$/H$_2$O dispersions exhibit much longer lifetimes compared to ME samples. The long lived species on a timescale of nanoseconds may be due to thermal repopulation of small-sized multilayer A-exctions acting as traps. For both the LPE 10-30k $g$ WS$_2$/H$_2$O and WS$_2$/IPA dispersions, $A_{FL}$-exciton GSB rises simultaneously while $A_{ML}$-exciton GSB goes through a fast decay at a timescale of picoseconds. The decay of the $A_{ML}$-exciton GSB and the initial rise and later decay of the $A_{FL}$-exciton GSB are fitted simultaneously with the constraint that the initial decay constant of the $A_{ML}$-exciton GSB and rise of the $A_{FL}$-exciton GSB are the same. Satisfactory fits are obtained with three exponential decays and an additional initial exponential rise for the $A_{FL}$-exciton GSB decays. The concomitant rise and decay of mono and multilayer signals indicates that there is energy transfer between monolayers and multilayers in LPE WS$_2$ dispersions, which can be responsible for the low PLQE in high quality monolayer enriched WS$_2$ dispersions prepared by liquid exfoliation.

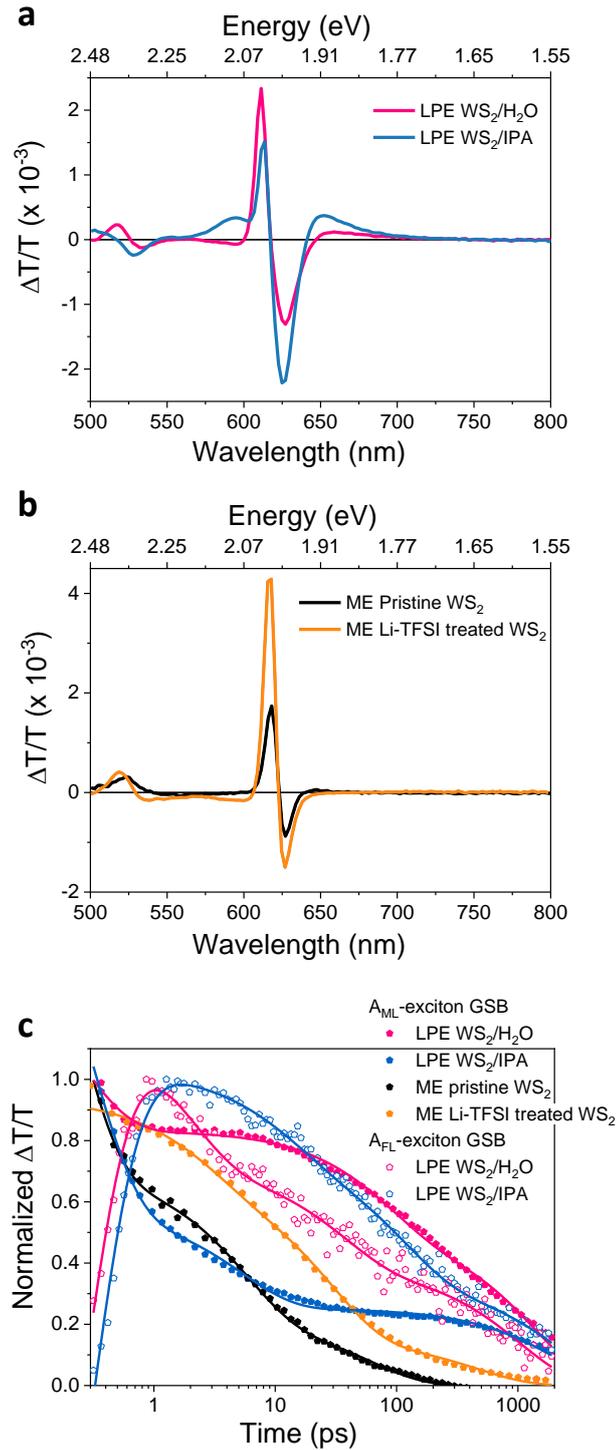

Figure 3. Ultrafast exciton dynamics. Pump-probe spectra at 10 ps time delay (610 nm excitation and 2.63 nJ/pulse) of **a** LPE 10-30k *g* WS$_2$/H$_2$O and LPE 10-30k *g* WS$_2$/IPA samples, and **b** ME pristine and Li-TFSI treated WS$_2$ monolayer samples. **c** Normalized kinetics taken at the A$_{ML}$-exciton GSB and A$_{FL}$-exciton GSB. Data are well fitted by a three exponential function (solid lines).

To test our hypothesis, we conduct further pump-probe measurement by exciting the LPE 10-30k *g* WS$_2$/IPA dispersion sample with an energy below the WS$_2$ bandgap, at 650 nm, to directly excite the multilayer components and observe the exciton decay. Since the position of $A_{FL}$-exciton GSB redshifts with increasing layer number, all the positive features shown in Figure 4a are assigned to the $A_{FL}$-exciton GSB. As summarized in Table S3, simultaneous with the fast $A_{FL}$-excition (~ 620 nm) decay is the growth of more aggregated $A_{FL}$-excition (~ 660 nm) GSB. The further extended $<\tau>$ at $A_{FL}$-exciton GSB (~ 660 nm) supports our assumption that there is energy transfer between the individual sheets in the LPE $WS_2$ dispersion. In this scenario, we propose that the optical quality of the LPE $WS_2$ can potentially be improved by reducing the energy transfer by introducing a coating on the nanosheets, for example through chemical functionalization or adsorption of bulky molecules or polymers.

In order to gain further insight into exciton dynamics of the $WS_2$ samples, we also analyse the photon energy dependence and pump fluence dependence of the A-exciton GSB decay feature (Figure S6-S9; Table S4-S7). A sub-picosecond decay component in the excited-state dynamics of $WS_2$ emerges for incident photon energies above the A-exciton resonance. This originates from a nonequilibrium population of charge carriers that form excitons as they cool, and is dependent on the photon energy.[41] Nevertheless, the exciton decay for LPE $WS_2$ samples and ME pristine $WS_2$ samples are largely independent on the pump fluence (Figure S7 and S8). The fluence independent nature of the recombination indicates that it is linked to defect-assisted decay.[42] While the $<\tau>$ at $A_{ML}$-exciton GSB for ME Li-TFSI treated $WS_2$ sampled shortens with the increase of pump fluence due to the enhanced EEA process. This suggests that there are also other reasons for the low PLQE of LPE $WS_2$ dispersions besides energy transfer, such as edge effects related to the small lateral dimensions of the monolayers in LPE dispersions. Hence the liquid exfoliation process needs to be further improved to produce samples suitable for practical optoelectronic application.[30]

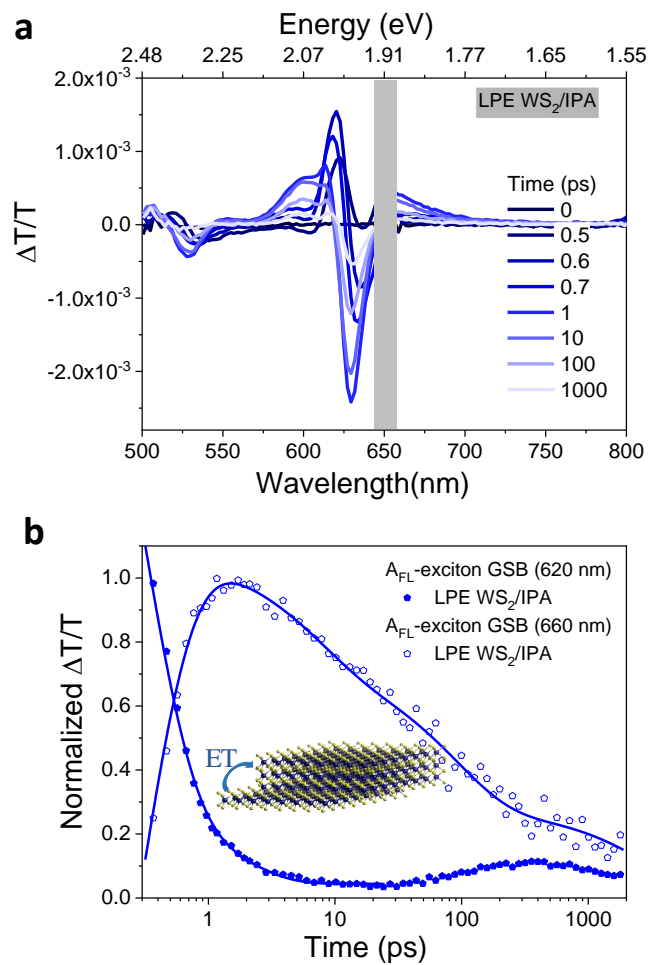

Figure 4. **a** Pump-probe spectra of the LPE 10-30k *g* WS$_2$/IPA sample at 650 nm excitation with 5.26 nJ/pulse. **b** Normalized kinetics taken at A$_{ML}$-exciton GSB and A$_{FL}$-exciton GSB, illustrating energy transfer (ET) from monolayers to multilayers. Data are well fit by three exponential decay function and an additional rise component for the A$_{FL}$-exciton GSB (solid lines).

In conclusion, monolayer enriched WS$_2$ dispersions produced by liquid phase exfoliation combined with size selection and WS$_2$ monolayers by mechanical exfoliation were prepared and the quality of the samples compared and investigated with respect to the optical and photophysical properties. Our results reveal that LPE monolayer-enriched WS$_2$ dispersions show pristine-like excitonic feature with narrow linewidth PL and minimal Stoke shift. In addition, we conduct a detailed analysis on the exciton dynamics of LPE WS$_2$ dispersions and ME WS$_2$ monolayers through pump-probe spectroscopy, and elucidate that there is an energy transfer between individual monolayer and multilayers nanosheets in the LPE WS$_2$ dispersions which is at least partly responsible for their low PLQE in addition to other factors such as small lateral sheet size. We also propose that a thicker coating shielding the sheets such as polymers as additional stabilizers can potentially reduce this phenomenon. As such, the study clearly identifies anchor points for further improving the liquid exfoliation methodology to produce materials with great potential in practical optoelectronic applications.

**Experimental Methods**

**Material**

WS$_2$ powder (99%, 2 μm, Sigma A), surfactant sodium cholate hydrate (SC, ≥ 99%), and bis(trifluuoromethane)sulfonimide lithium salt (Li-TFSI) are purchased from Sigma-Aldrich and used without purification. The bulk synthetic WS$_2$ crystal is purchased from 2D Semiconductors. The mechanically-exfoliated monolayer WS$_2$ is prepared according to the reported gold-mediated exfoliation method to ensure relatively large monolayers.[43]

**Liquid exfoliation process**

WS$_2$ powder (30 g/L) and SC (8 g/L) are added in a glass bottle with 80 mL deionized water and the dispersion is transferred to a stainless-steel beaker for sonication. The beaker is placed in a cooled water bath with a temperature of 5 °C (maintained through a chiller). An ultrasonic replaceable tip is positioned in the dispersion ~2 cm from the bottom and the mixture is sonicated for 1 h with 60% amplitude (pulse 8 s on and 2 s off ratio), using a Sonics Vibracell VCX 500, equipped with a threaded probe. The metal beaker is covered with aluminium foil during the sonication process. After the sonication, the dispersion is centrifuged at 6000 RPM for 1.5 h at 8 °C in a Hettich Mikro 220R centrifuge, equipped with a 1016 fixed-angle rotor. The participants are removed afterwards and 2 g/L SC solution is added to the dispersion to reach 80 mL, followed by another tip sonication with same amplitude at 5 °C for 5.5 h. The first sonication step serves the purpose of removing impurities in the WS$_2$ powder. After the second sonication, the dispersion is transferred to centrifuge tubes for size selection by liquid cascade centrifugation.[20] First, the dispersions are centrifuged with relative centrifugal force (RCF) 5k *g* for 1 h at 8 °C. Supernatant and sediment are separated through manual pipetting. The supernatant is collected and centrifuged with the same speed for 2 h at 8 °C to remove large/thick sheets as completely as possible. Then the supernatant is centrifuged with RCF 10k *g* for 2 h at 8 °C. The sediment is collected and dispersed in 0.1 g/L SC solution (~ 2mL), which is referred to 5-10k *g* WS$_2$/H$_2$O sample. The supernatant is transferred to new centrifuge tubes for further centrifugation with RCF 30k *g* for 2 h at 8 °C. In the end, the supernatant is discarded, while the sediment is collected and dispersed in 0.1 g/L SC solution (~ 2mL), which is referred to 10-30k *g* WS$_2$/H$_2$O sample. To transfer the 10-30k *g* WS$_2$/H$_2$O sample from water to IPA, the dispersion is centrifuged with RCF 30k *g* for 1.5 h to pellet out the nanosheets as sediment and decant the water supernatant. The sediment is redispersed in IPA through 5 min bath sonication. Then the dispersion is centrifuged with RCF 2k $g$ for 20 min to remove the majority of aggregates. The supernatant is collected as 10-30k $g$ WS$_2$/IPA sample.

**Chemical treatment**

The chemical treatment with Li-TFSI (0.02 M in Methanol) is carried out in ambient atmosphere. The chemical treatments are achieved by immersing the samples into concentrated solutions of the investigated chemicals for 40 mins, and blow dry with nitrogen gun afterwards.

**Supporting Information**

Supporting Information is available online with additional experimental details as well as additional data for optical and photophysical characterization of WS$_2$ samples.

Data available in University of Cambridge data repository at: link to be added during proof.

**Acknowledgement**

This project has received funding from the European Research Council (ERC) under the European Union's Horizon 2020 research and innovation program (Grant Agreement No…..). Z.L. acknowledges funding from the Swedish research council, Vetenskapsrådet 2018-06610. We acknowledge financial support from the EPSRC and the Winton Programme for the Physics of Sustainability. C. B. acknowledges support from the German research foundation (DFG) under grant agreement Emmy-Noether, BA4856/2-1 and Jana Zaumseil for the access to the infrastructure at the Chair of Applied Physical Chemistry.

Supporting Information

# Photophysical Comparison of Liquid and Mechanically Exfoliated WS$_2$ Monolayers

Zhaojun Li[1,2], Farnia Rashvand[3], Hope Bretscher[1], Beata M. Szydłowska[3], James Xiao[1], Claudia Backes[3,4], Akshay Rao[1*]

[1]Cavendish Laboratory, University of Cambridge, JJ Thomson Avenue, CB3 0HE, Cambridge, United Kingdom

[2]Molecular and Condensed Matter Physics, Department of Physics and Astronomy, Uppsala University, 75120 Uppsala, Sweden

[3]Institute for Physical Chemistry, Ruprecht-Karls-Universität Heidelberg, Im Neuenheimer Feld 253, 69120 Heidelberg, Germany

[4]Current address: Physical Chemistry of Nanomaterials, University of Kassel, Heinrich-Plett-Str.40, D-34132 Kassel, Germany

[*]Corresponding author

Contact e-mail: ar525@cam.ac.uk

## Contents

# 1. Experimental Details

Si/SiO$_2$ substrates with 90 nm oxide layer were used for microscope steady-state photoluminescence (PL) and Raman spectroscopy. Quartz substrates were used for ultrafast pump-probe measurement. The samples were encapsulated for ultrafast pump-probe measurements, and other measurements are carried out on samples without encapsulation.

Transmission electron microscopy (TEM) was performed using a FEI Tecnai F20 at 200kV accelerating voltage. 10-30k *g* WS$_2$/H$_2$O dispersion sample was transferred onto 200-mesh Cu grids (Agar AGS160). The optical microscopy was measured using Olympus BX60 optical microscope with 405 nm laser. Samples were placed on an X-Y piezo stage of the microscope and the signal is collected in refection mode with the 50× objective. The Raman spectroscopy was carried out on a Renishaw inVia Raman confocal microscope with a 532 nm excitation laser in air under ambient condition. The Raman emission was collected by a 20× long working distance objective lens in streamline mode and dispersed by a 1800 l/mm grating with 1% of the laser power (< 10 μW). The spectrometer was calibrated to a silicon reference sample prior to the measurement to correct for the instrument response.

UV-Vis measurement was performed using a Shimadzu UV-3600 Plus spectrometer to measure the extinction spectra of the LPE WS$_2$ dispersions in transmission mode in quartz cuvettes with 1 nm increments. Steady-state PL measurement were carried out using a temperature and current-controlled 405 nm laser diode (Thorlabs). The incident beam was attenuated as desired and focused onto the sample while PL from the sample was collected and focused into an Andor Kymera 328i Spectrometer and spectra recorded using a Si-CCd (Andor iDus 420).

The microscope steady-state PL measurement was carried out using a WITec alpha 300 s setup as has been described previously.[1] Importantly, a 405 nm continuous wave laser (Coherent CUBE) was used as the excitation source. A long pass filter with a cutoff wavelength of 450 mm was fitted before signal collection to block excitation scatter. The light was coupled with an optical fiber to the microscope and focused using a 20× Olympus lens. Samples were placed on an X-Y piezo stage of the microscope. The PL signal was collected in refection mode with the same 20× objective and detected using a Princeton Instruments SP-2300i spectrometer fitted with an Andor iDus 401 CCD detector. The PL was measured at 405 nm excitation with a fluence of 15 W cm$^{-2}$.

The ultrafast pump-probe setup has been described previously.[2] A Light Conversion PHAROS laser system with 400 µJ per pulse at 1030 nm with a repetition rate of 38 kHz is split in two, one part is used to generate the continuum probe light and the second part is used in an Collinear Optical Parametric Amplifier (Orpheus, Light Conversion) to generate the pump source at the desired wavelength. The probe pulse is delayed up to 2 ns with a mechanical delay-stage (Newport). A mechanical chopper (Thorlabs) is used to create an on-off pump-probe pulse series. A silicon line scan camera (JAI SW-2000M-CL-80) fitted onto a visible spectrograph (Andor Solis, Shamrock) is used to record the transmitted probe light.

## 2. Calculation of Monolayer Content

The monolayer volume fraction ($V_f$) can be extracted from the extinction spectra of the dispersion according to the previously reported method.[3] The A-exciton extinction is first deconvoluted into components of monolayered and multi-layered WS$_2$ with differentiation after smoothing the spectra by the Lowess method (10-15 points). The smoothing function suppresses spectral noise well without changing peak shapes. In the simplest form, a Lorentzian line can be described by

$$L(E) = \frac{h}{\left[1 + \left(\frac{(E-E_0)}{w/2}\right)^2\right]} \quad (S1)$$

Where $h$ represents the height, $E_0$ the center and $w$ the full width half maximum (FWHM). Differentiating twice with respect to $E$ gives

$$\frac{d^2L(E)}{dE^2} = -\frac{8h}{w^2}\left[\frac{1 - 3\left(\frac{E-E_0}{w/2}\right)^2}{\left(1 + \left(\frac{E-E_0}{w/2}\right)^2\right)^3}\right] \quad (S2)$$

The obtained spectrum of the second derivative is then fitted to the sum of the second derivative of two Lorentzian functions giving $E$, $w$ and $h$ of the monolayer and few-layer WS$_2$. The area under the monolayer (ML) A-exciton extinction peak should scale with the monolayer content in the dispersion. As the area under any Lorentzian is proportional to $h \times w$, a metric $S_A$ which scales with the $V_f$ can be calculated as the equation:

$$S_A = \frac{h_{ML}w_{ML}}{h_{ML}w_{ML} + h_{FL}w_{FL}} \tag{S3}$$

$V_f$ is then calculated as:[3]

$$V_f = (1.25 \pm 0.08)S_A \tag{S4}$$

## 3. Material characterization

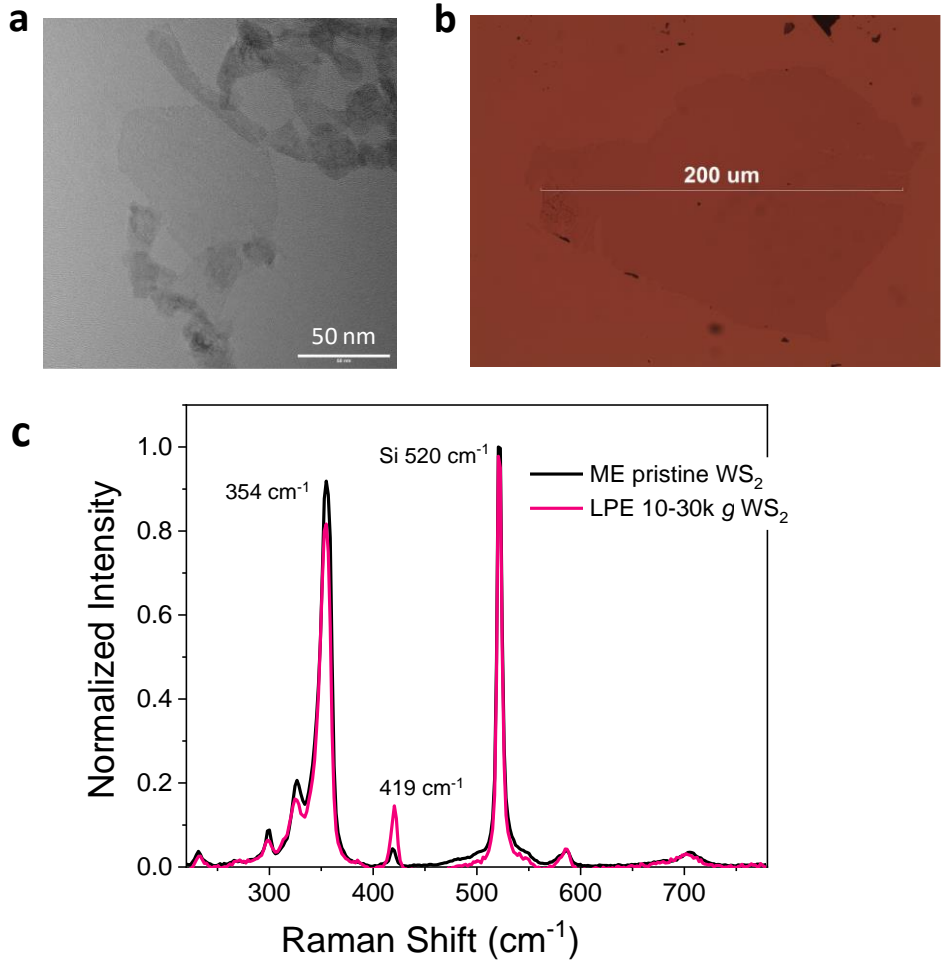

Figure S1. Characterization of liquid exfoliated and mechanically exfoliated $WS_2$ samples. **a** TEM image of liquid phase exfoliated 10-30k *g* $WS_2$ sample. **b** Optical microscope image of mechanically exfoliated $WS_2$ sample on quartz substrate. **c** Raman spectroscopy of liquid phase exfoliated and mechanically exfoliated $WS_2$ samples on $Si/SiO_2$ substrate confirming the monolayer.

## 4. Optical properties of LPE WS$_2$ dispersions and ME WS$_2$ samples

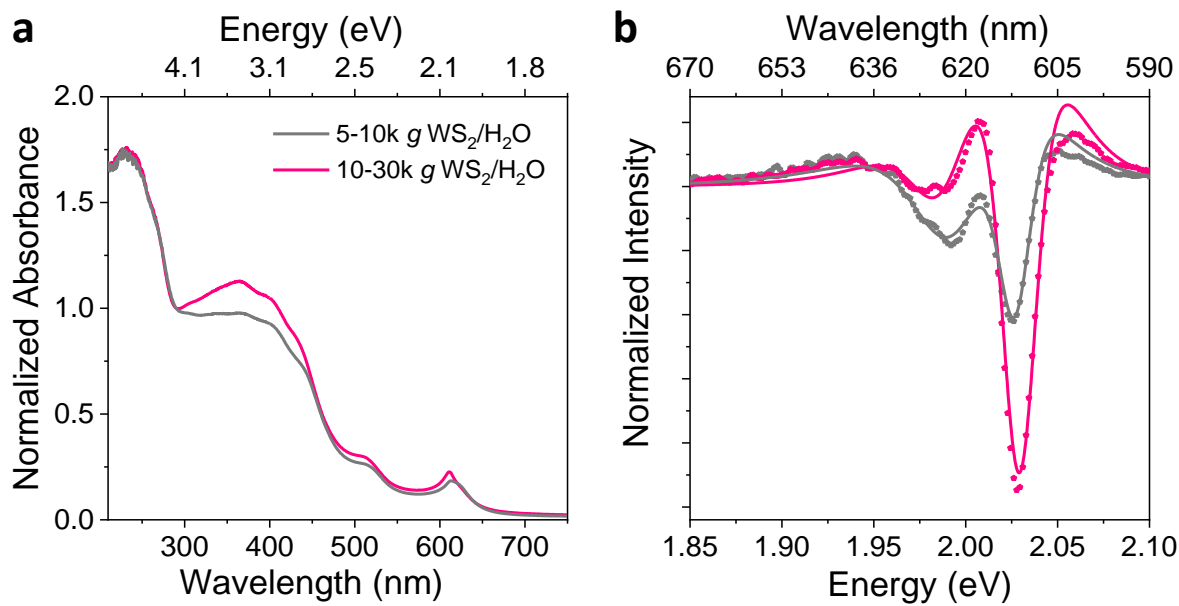

Figure S2. **a** Extinction spectra normalized to 290 nm of the 5-10k *g* and 10-30k *g* WS$_2$/H$_2$O samples. **b** Second derivatives of the A-exciton obtained after smoothing the spectrum with the Lowess method. The spectra are fitted to the second derivative of two Lorentzians (solid lines).

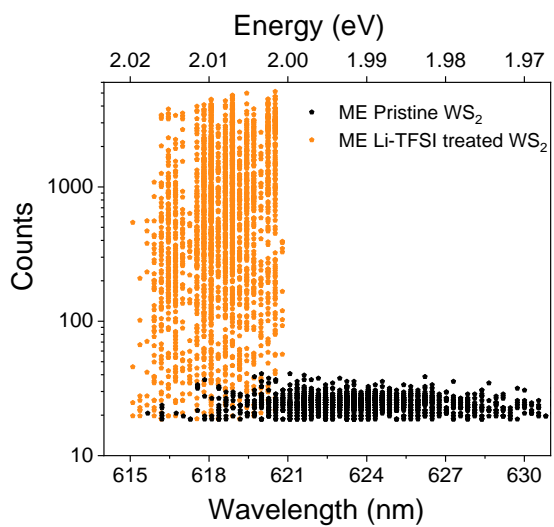

Figure S3. PL scatter plots of spectral position of the peak emission as well as peak pristine and Li-TFSI-treated monolayer WS$_2$ PL counts extracted from PL maps of WS$_2$ monolayer on Si-SiO$_2$ (90 nm) substrate.

# 5. Additional Pump-probe data on WS$_2$ samples

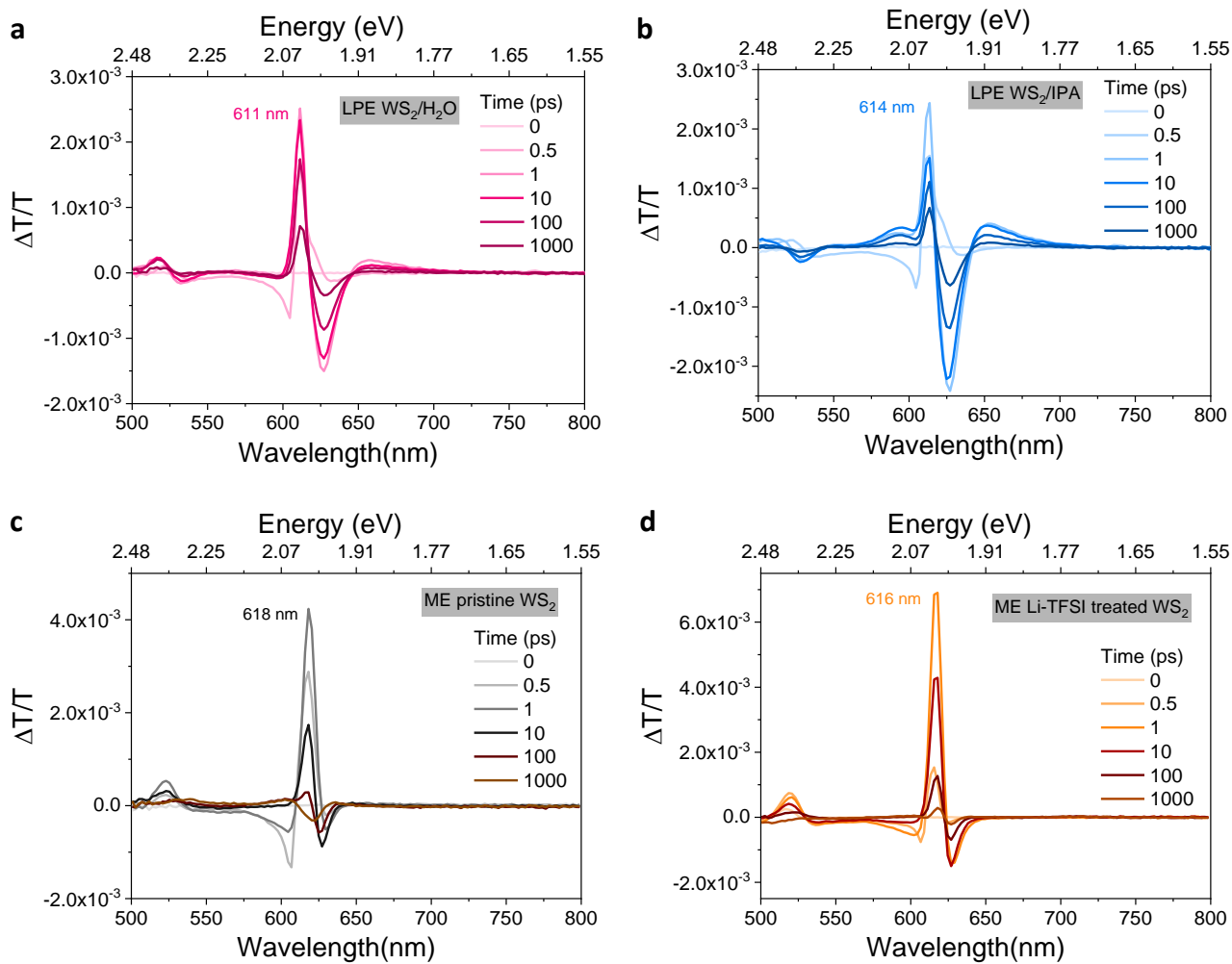

Figure S4. Pump-probe spectra (excited at 610 nm, 2.63 nJ/pulse) of **a** liquid exfoliated 10-30k *g* WS$_2$/H$_2$O, **b** liquid phase exfoliated 10-30k *g* WS$_2$/IPA, **c** mechanically exfoliated pristine WS$_2$ monolayer sample, and **d** mechanically exfoliated Li-TFSI treated WS$_2$ monolayer sample.

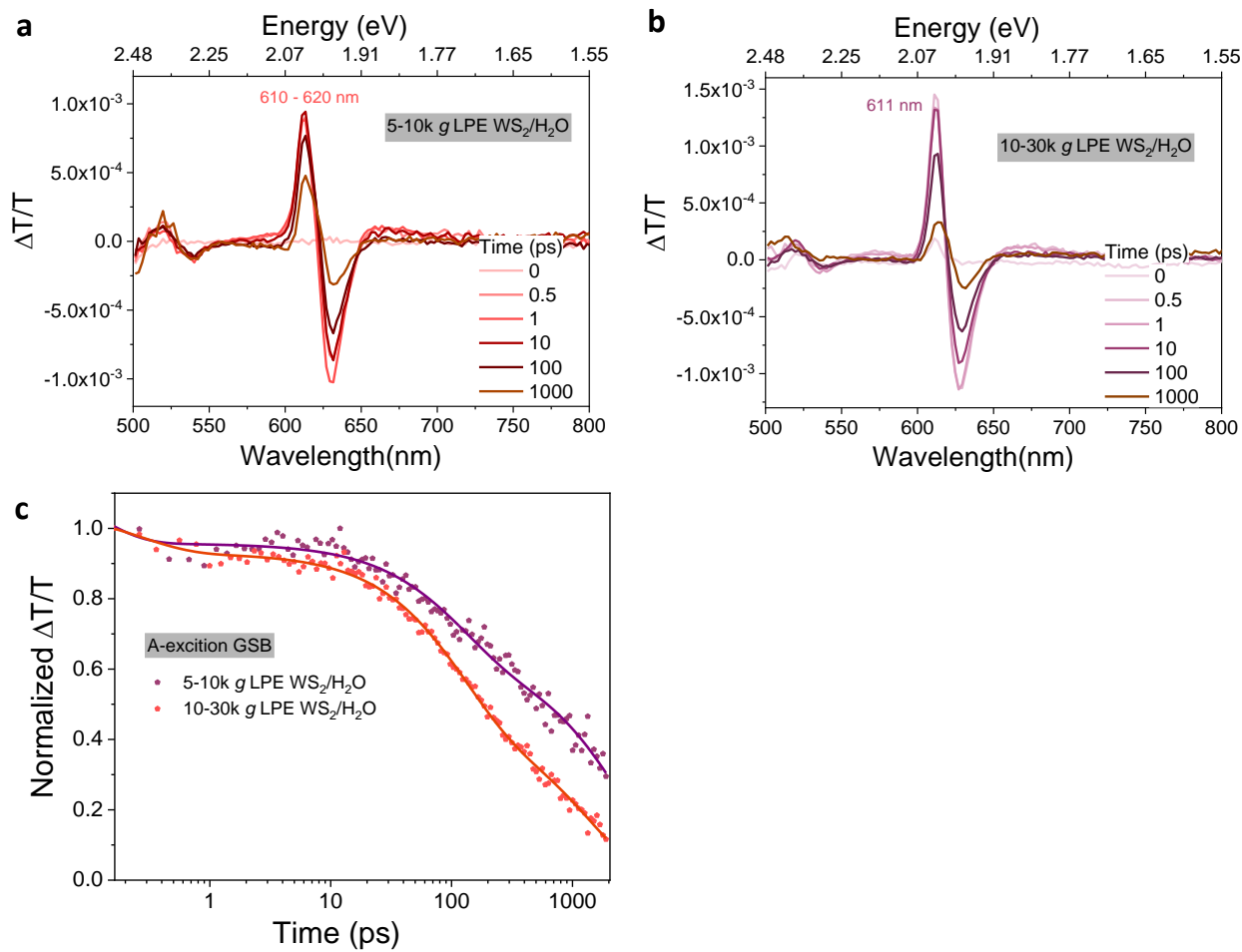

Figure S5. Pump-probe spectra (excited at 500 nm, 0.53 nJ/pulse) of **a** liquid phase exfoliated 5-10k *g* WS$_2$/H$_2$O sample, **b** liquid phase exfoliated 10-30k *g* WS$_2$/ H$_2$O sample. c Normalized kinetics of A-exciton GSB of both samples.

Table S1. Fitting results of the rates for LPE 10-30k *g* WS$_2$/IPA sample at 500 nm excitation with 0.53 nJ/pulse in pump-probe measurement.

| Sample | A$_1$ | τ$_1$ (ps) | A$_2$ | τ$_2$ (ps) | A$_3$ | τ$_3$ (ps) | <τ> (ps) |
|---|---|---|---|---|---|---|---|
| LPE 5-10k *g* WS$_2$/H$_2$O A$_{ML}$-exciton GSB (~611 nm) | 0.25 | 0.10 | 0.32 | 114 | 0.63 | 2591 | 1391 |
| LPE 10-30k *g* WS$_2$/ H$_2$O A-exciton GSB (610~620 nm) | 0.13 | 0.26 | 0.48 | 116 | 0.45 | 1419 | 655 |

Compared to the LPE 10-30k *g* WS$_2$/IPA dispersion, both 5-10k *g* and 10-30k *g* WS$_2$/H$_2$O show less positive features in the 650 − 680 nm region in the pump-probe spectra, suggesting the fraction of multilayers are less than that of WS$_2$/IPA samples (Figure S4 and S5). This is understandable for 10-30k *g* WS$_2$/H$_2$O sample since the V$_f$ is really high, around 78%, however, the V$_f$ in 5-10k *g* WS$_2$/H$_2$O sample is quite low (~ 15%). On the other hand, compared to the narrow linewidth of A$_{ML}$-exciton GSB for both LPE 10-30k *g* WS$_2$/H$_2$O and WS$_2$/IPA, the 5-10k *g* sample shows wider linewidth and even peak splitting in the longer time delay stage. In addition, the A$_{ML}$-exciton GSB of the 5-10k *g* sample decays slower than that LPE 10-30k *g* WS$_2$/H$_2$O (Figure S5 c and Table S1), indicating there are multilayers which are not distinguishable from monolayer in 5-10k *g* sample and possible energy transfer between them. Therefore, we assume the multilayers existing in 5-10k *g* WS$_2$/H$_2$O and 10-30k *g* WS$_2$/IPA are different. There are more double or triple layers in 5-10k *g* WS$_2$/H$_2$O sample which is removed with high centrifugation speed, while there are more aggregated multilayers in 10-30k *g* WS$_2$/IPA sample due to the reaggregation occurring during the solvent exchange process.

Table S2. Fitting results for the rates at 610 nm excitation with 2.63 nJ/pulse in pump-probe measurement.

| Sample | A$_1$ | $\tau_1$ (ps) | A$_2$ | $\tau_2$ (ps) | A$_3$ | $\tau_3$ (ps) | A$_4$ | $\tau_4$ (ps) | $<\tau>$ (ps) |
|---|---|---|---|---|---|---|---|---|---|
| LPE 10-30k *g* WS$_2$/H$_2$O A$_{ML}$-exciton GSB | 0.65 | 0.23 | 0.34 | 82 | 0.49 | 1396 | / | / | 481 |
| LPE 10-30k *g* WS$_2$/H$_2$O A$_{FL}$-exciton GSB | -3.6 | 0.23 | 0.54 | 1.8 | 0.33 | 35 | 0.38 | 1034 | 360 |
| LPE 10-30k *g* WS$_2$/IPA A$_{ML}$-exciton GSB | 1.64 | 0.26 | 0.32 | 5.3 | 0.25 | 2032 | / | / | 231 |
| LPE 10-30k *g* WS$_2$/IPA A$_{FL}$-exciton GSB | -3.6 | 0.26 | 0.26 | 12 | 0.43 | 111 | 0.34 | 2151 | 759 |
| ME pristine WS$_2$ A$_{ML}$-exciton GSB | 0.74 | 0.18 | 0.18 | 5.5 | 0.08 | 63 | / | / | 6 |
| ME Li-TFSI treated WS$_2$ A$_{ML}$-exciton GSB | 0.29 | 2.7 | 0.55 | 29 | 0.16 | 466 | / | / | 90 |

Table S3. Fitting results for the rates for LPE 10-30k *g* WS$_2$/IPA sample at 650 nm excitation with 5.26 nJ/pulse in pump-probe measurement.

| Sample | $A_1$ | $\tau_1$ (ps) | $A_2$ | $\tau_2$ (ps) | $A_3$ | $\tau_3$ (ps) | $A_4$ | $\tau_4$ (ps) | $\langle\tau\rangle$ (ps) |
|---|---|---|---|---|---|---|---|---|---|
| LPE 10-30k $g$ WS$_2$/IPA $A_{FL}$-exciton GSB (~620 nm) | 2.64 | 0.28 | 0.27 | 1.5 | 0.04 | 367 | / | / | 6 |
| LPE 10-30k $g$ WS$_2$/IPA $A_{FL}$-exciton GSB (~660 nm) | -2.95 | 0.28 | 0.35 | 6.7 | 0.45 | 89 | 0.27 | 3156 | 836 |

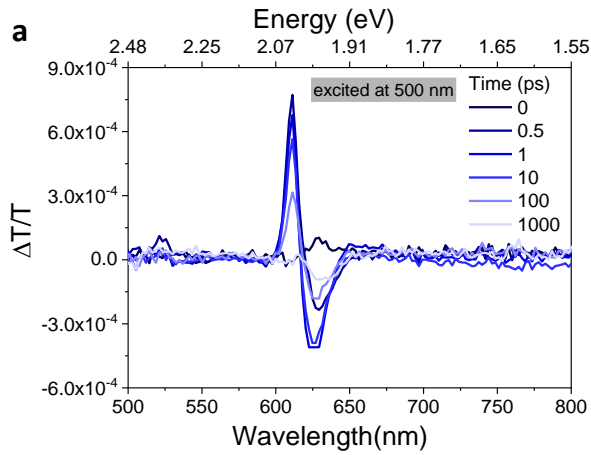
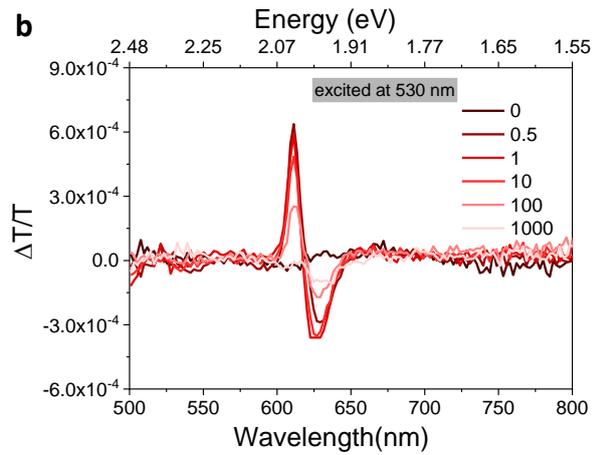
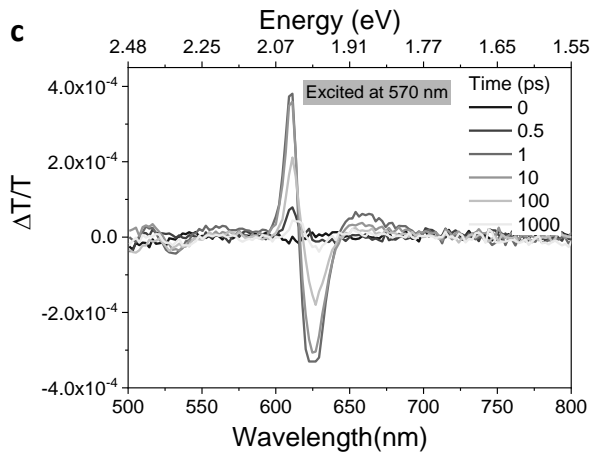
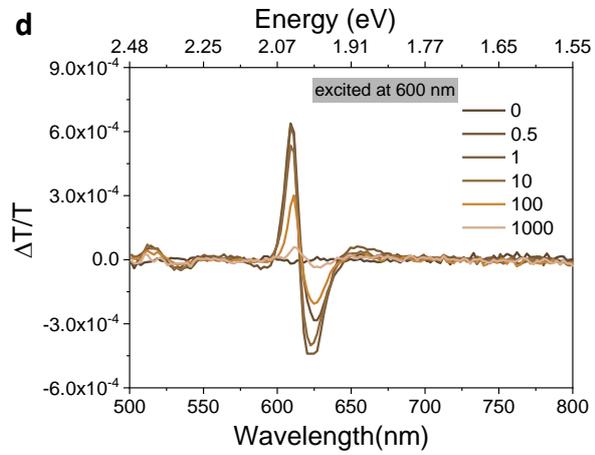
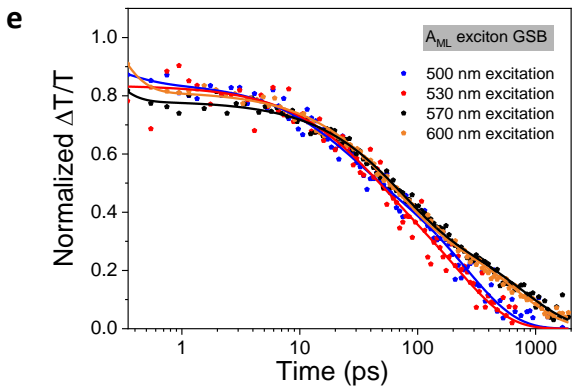

Figure S6. Photon energy dependence of A-exciton resonance with 0.53 nJ/pulse in pump-probe measurements. **a** pump-probe data of the 10-30k *g* WS$_2$/H$_2$O sample excited at **a** 500 nm, **b** 530 nm, **c** 570 nm, and **d** 600 nm. **e** Normalized kinetics of A$_{ML}$-exciton GSB of all samples.

Table S4. Fitting results for the rates of the 10-30k *g* WS$_2$/H$_2$O sample with 0.53 nJ/pulse in pump-probe measurement.

| Sample | A$_1$ | $\tau_1$ (ps) | A$_2$ | $\tau_2$ (ps) | A$_3$ | $\tau_3$ (ps) | $\langle\tau\rangle$ (ps) |
|---|---|---|---|---|---|---|---|
| 500 nm excitation | 0.16 | 0.2 | 0.28 | 21.7 | 0.57 | 254 | 149 |
| 530 nm excitation | 0.15 | 0.07 | 0.29 | 30.5 | 0.54 | 223 | 132 |
| 570 nm excitation | 1.2 | 0.1 | 0.42 | 65.2 | 0.36 | 763 | 153 |
| 600 nm excitation | 1.4 | 0.1 | 0.44 | 65.1 | 0.38 | 655 | 125 |

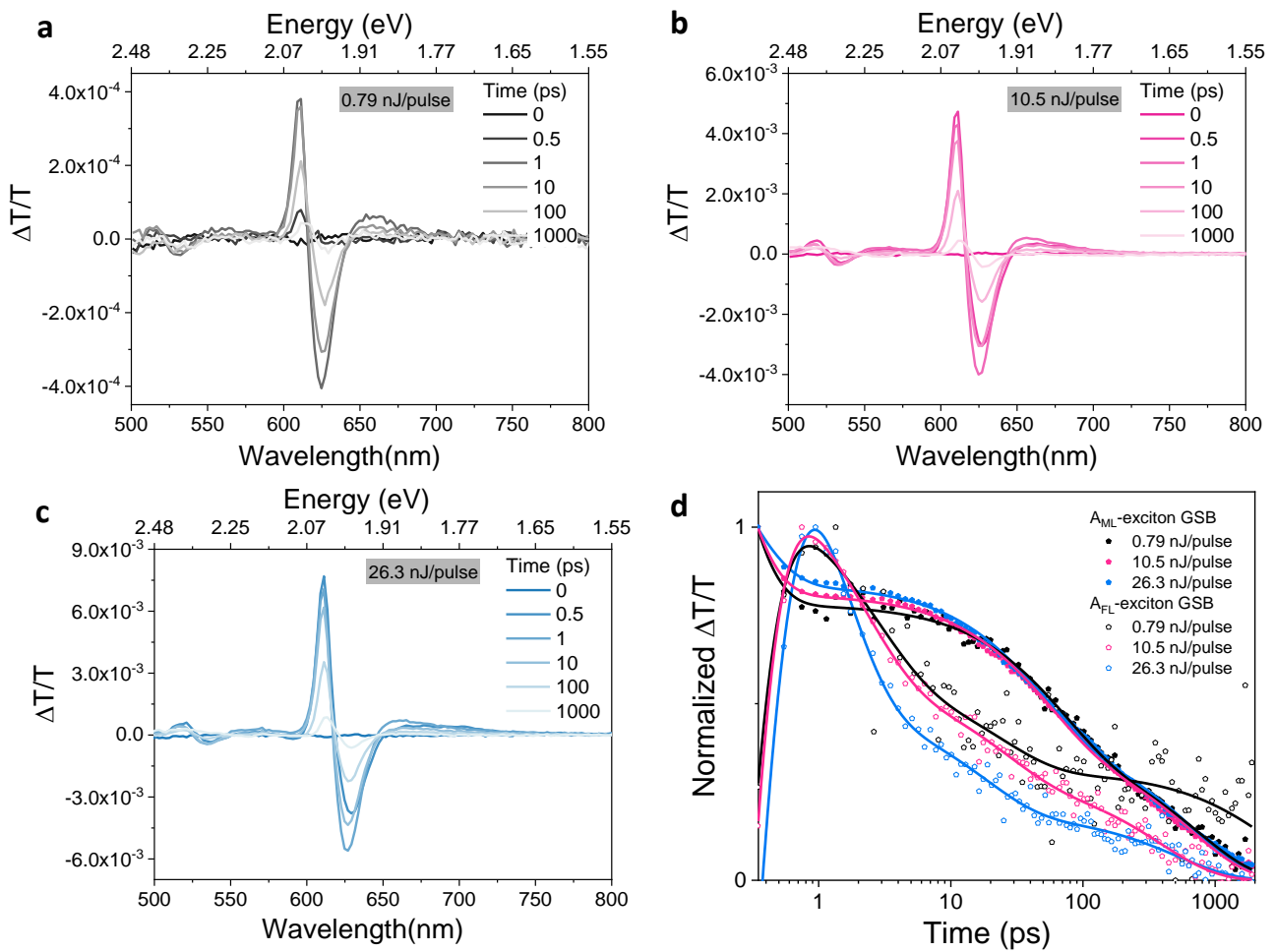

Figure S7. Fluence dependence of A-exciton resonance (excited at 570 nm). **a** pump-probe data of the 10-30k *g* $WS_2/H_2O$ sample excited with **a** 0.79 nJ/pulse, **b** 10.5 nJ/pulse, and **c** 26.3 nJ/pulse. **d** Normalized kinetics of $A_{ML}$-exciton GSB and $A_{FL}$-exciton GSB of all samples.

Table S5. Fitting results for the rates at 570 nm excitation of the 10-30k *g* $WS_2/H_2O$ sample in pump-probe measurement.

| Sample | $A_1$ | $\tau_1$ (ps) | $A_2$ | $\tau_2$ (ps) | $A_3$ | $\tau_3$ (ps) | $A_4$ | $\tau_4$ (ps) | $<\tau>$ (ps) |
|---|---|---|---|---|---|---|---|---|---|
| 0.79 nJ/pulse $A_{ML}$-exciton GSB | 2.1 | 0.14 | 0.42 | 66 | 0.36 | 766 | / | / | 106 |
| 0.79 nJ/pulse $A_{FL}$-exciton GSB | -8.9 | 0.14 | 0.57 | 2.6 | 0.27 | 24 | 0.31 | 2692 | 733 |
| 10.5 nJ/pulse $A_{ML}$-exciton GSB | 1.97 | 0.15 | 0.43 | 51 | 0.38 | 667 | / | / | 99 |
| 10.5 nJ/pulse $A_{FL}$-exciton GSB | -10.2 | 0.15 | 0.69 | 2.0 | 0.28 | 24 | 0.28 | 386 | 93 |
| 26.3 nJ/pulse $A_{ML}$-exciton GSB | 0.8 | 0.2 | 0.44 | 47 | 0.4 | 685 | / | / | 180 |
| 26.3 nJ/pulse $A_{FL}$-exciton GSB | -8.29 | 0.2 | 1.32 | 1.3 | 0.3 | 17.6 | 0.19 | 518 | 58 |

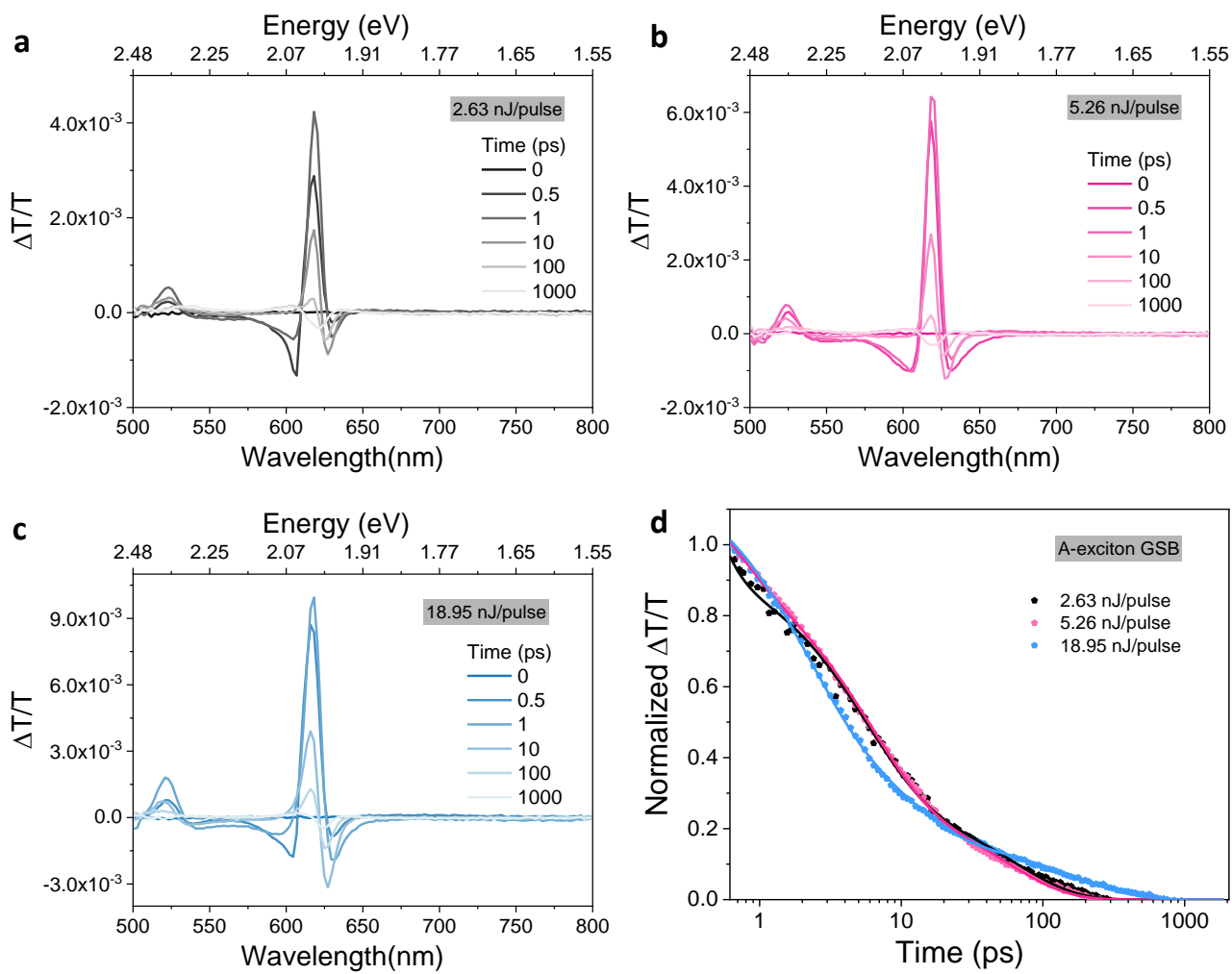

Figure S8. Fluence dependence of A-exciton resonance (excited at 610 nm) in pump-probe measurements. Pump-probe data of mechanically exfoliated pristine monolayer $WS_2$ sample with **a** 2.63 nJ/pulse, **b** 5.26 nJ/pulse, and **c** 18.95 nJ/pulse. **d** Normalized kinetics of A-exciton GSB of all samples.

Table S6. Fitting results for the rates at 610 nm excitation of mechanically exfoliated pristine $WS_2$ monolayer sample in pump-probe measurement.

| Sample | $A_1$ | $\tau_1$ (ps) | $A_2$ | $\tau_2$ (ps) | $A_3$ | $\tau_3$ (ps) | $<\tau>$ (ps) |
|---|---|---|---|---|---|---|---|
| 2.63 nJ/pulse | 2.68 | 0.18 | 0.67 | 5.5 | 0.29 | 62 | 6 |
| 5.62 nJ/pulse | 0.44 | 0.47 | 0.68 | 5.7 | 0.29 | 55 | 14 |
| 18.95 nJ/pulse | 0.68 | 1.93 | 0.37 | 10.0 | 0.18 | 167 | 29 |

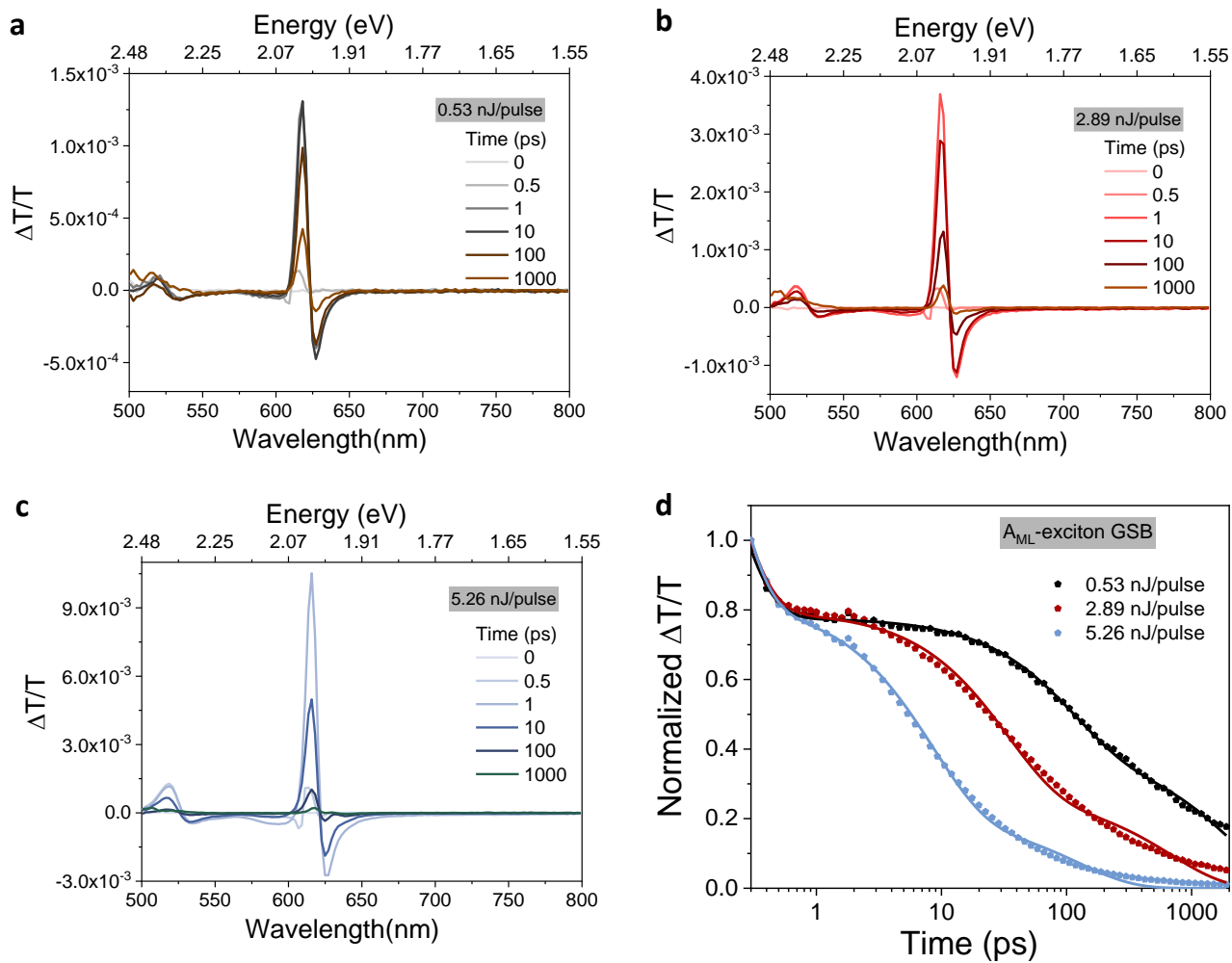

Figure S9. Fluence dependence of A-exciton resonance (excited at 610 nm) in mechanically exfoliated Li-TFSI treated monolayer WS$_2$ sample. Pump-probe data with **a** 0.53 nJ/pulse and **b** 2.83 nJ/pulse. **c** 5.26 nJ/pulse **d** Normalized kinetics taken at A$_{ML}$-exciton GSB of all samples.

Table S7. Fitting results for the rates at 610 nm excitation of mechanically exfoliated Li-TFSI treated WS$_2$ monolayer sample in pump-probe measurement.

| Sample | $A_1$ | $\tau_1$ (ps) | $A_2$ | $\tau_2$ (ps) | $A_3$ | $\tau_3$ (ps) | $\langle\tau\rangle$ (ps) |
|---|---|---|---|---|---|---|---|
| 0.53 nJ/pulse | 1.77 | 0.14 | 0.39 | 108 | 0.39 | 2024 | 324 |
| 2.83 nJ/pulse | 2.05 | 0.13 | 0.26 | 688 | 0.54 | 32.8 | 70 |
| 5.26 nJ/pulse | 2.62 | 0.12 | 0.62 | 8 | 0.20 | 126 | 9 |